\documentclass[10pt.]{article}

\usepackage{amsmath}
\usepackage{amssymb}

\usepackage{graphicx}

\usepackage{cite}

\usepackage{color}

\usepackage[labelfont=bf,labelsep=period,justification=raggedright]{caption}

\makeatletter
\renewcommand{\@biblabel}[1]{\quad#1.}
\makeatother

\date{}

\begin{document}
\font\myfont=cmr12 at 25pt

\vspace*{0.2in}
\begin{flushleft}
{\Large
\bf{\myfont Glassy states of aging social networks  }
}
\bigskip
\\
\textbf{ F. Hassanibesheli$^{1,2 \diamondsuit }$, L. Hedayatifar$^{1,2 \diamondsuit }$, H. Safdari$^1$, M. Ausloos $^{3,4}$, G.~R. Jafari$^{1,5,6 \ast}$}
\newline
\\

\textbf{1} Department of Physics, Shahid Beheshti University, G.C., Evin, Tehran 19839, Iran\\
\textbf{2} AGH University of Science and Technology, Faculty of Physics and Applied Computer Science, al. Mickiewicza 30, 30-059 Krakow, Poland\\
\textbf{3} GRAPES, rue de la Belle Jardiniere 483, B-4031, Angleur, Belgium\\
%\textbf{4} eHumanities group,  Royal Netherlands Academy of Arts and Sciences, Joan Muyskenweg 25, 1096 CJ, Amsterdam, The Netherlands\\
\textbf{4} School of Business, University of Leicester, University Road, Leicester, LE1 7RH, United Kingdom\\
\textbf{5} The Institute for Brain and Cognitive Science (IBCS), Shahid Beheshti University,
 G.C., Evin, Tehran 19839, Iran\\
\textbf{6} Center for Network Science, Central European University, H-1051, Budapest, Hungary
\newline
\\

$\diamondsuit$  These authors contributed equally to this work.

$\ast$ E-mail: g\_jafari@sbu.ac.ir
\end{flushleft}

\date{\today}

\section*{Abstract}
Individuals often develop reluctance to change their social relations, called "secondary homebody", even though their interactions with their environment evolve with time. Some memory effect is loosely present deforcing changes. In other words, in presence of memory, relations do not change easily. In order to investigate some history or memory effect on social networks, we introduce a temporal kernel function into the Heider conventional balance theory, allowing for  the "quality" of past relations to contribute to the evolution of the system. This memory effect is shown to lead to the emergence of aged networks, thereby perfectly describing and the more so measuring the aging process of links ("social relations"). It is shown that such a memory does not change the dynamical attractors of the system, but does prolong the time necessary to reach the "balanced states". The general trend goes toward obtaining either global ("paradise" or "bipolar") or local ("jammed") balanced states, but is profoundly affected by aged relations. The resistance of elder links against changes decelerates the evolution of the system and traps it into so named $glassy$ $states$. In contrast to balance configurations which live on stable states, such long lived glassy states can survive in unstable states.

\section*{Introduction}
\textbf{''Yesterdays' friend (enemy) rarely become tomorrows' enemy (friend).''}\\

Tension reduction is a predominant principle that contributes to the formation of human interactions ~\cite{heider}. This principle acts as a self-organizing process; it indicates that social communications are established based on the tendency towards balanced states \cite{wang,perca,kirman,Ram}. Interesting questions that follow  concern what "parameters"  have a pivotal role in the social network dynamics. An appropriate answer seems to lie in the history of relationships. The ability of human beings to remember sequences of events (sometimes unconsciously) over the time brings about  social concepts, such as commitment and allegiance that lead to the formation of cultural communities, sects, alliances, and political groups ~\cite{psy1,psy2,psy3}. In psychological terms, the more potent the commitment is, the more probable the relations will remain unchanged over time. There are persons, such as family members, friends (or enemies), even business partners, with whom most people have no inclination to modify the nature of their relationships. This resistance to change can be explained by the history of relations and their importance which can be referred to two distinct parameters "age of links" and "weight of links", respectively. Generally, depending on age or weight, breaking up or modifying the links among individuals gradually can become difficult.

 In this study, the main intention is to explore the effect of relationships history following the hypothesis that  newly formed relations have more chance to change than older ones; thus  old relations are more resistant. Practically, we aim at mimicking the strength of relationship links through a model considering a combination of emotional intensity and life time duration of relations, but keeping these "parameters" as independent of each other as proposed by Granovetter~\cite{gra}. This leads to links which contain both aspects, - as depicted on Fig.  1  through different thicknesses of lines (for the strengths) and different colors (for the ages).

\begin{figure}
\centerline{\includegraphics[trim = 10mm 30mm 40mm 0mm, clip,width=.33\textwidth]{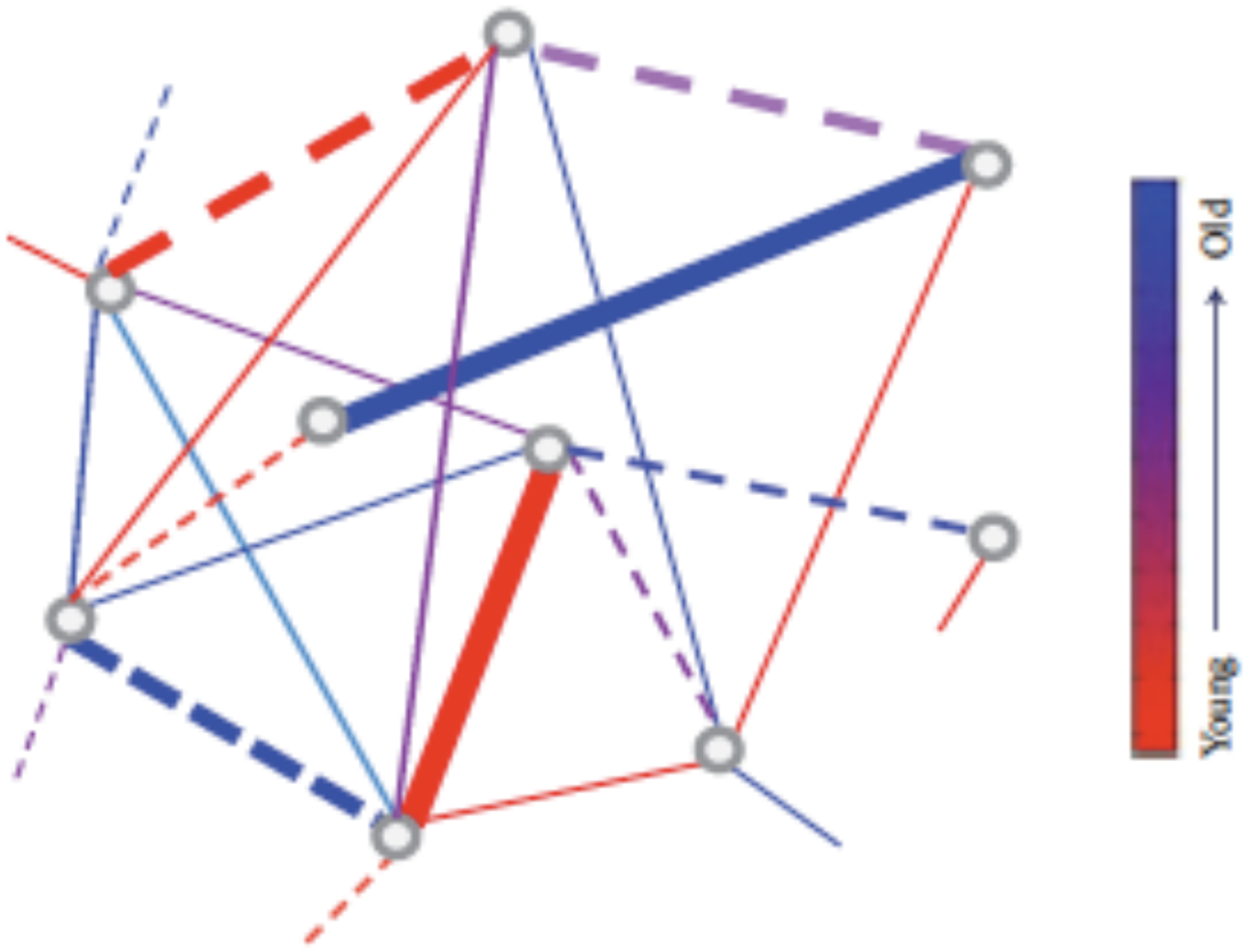}}
\caption{  In terms of social networks dynamics, links can carry various information such as type, age and strength of relations. This figure illustrates a network with two types of relations (for instance friendship and animosity) which are denoted by solid and dash lines respectively. Here colors display the gradient of age from young (red) to old (blue);  the weight (strength) of links are represented by the line thickness. Increasing the age or weight of links can lead to decreasing the tendency towards modifying relationships.}
\label{fig1}
\end{figure}

Therefore, in quantitative terms and in view of modeling the dynamics of social networks, it is useful to develop some study in order to comprehend how history (or memory) has global consequences on the evolution of a social system. Thanks to technological advances, allowing collection and analysis of empirical data about human interactions on social networks, it is nowadays easily revealed that social activities act as sequences of correlated events in which each event or decision depends on previous ones and on the "node environment". In this context, much effort has been already devoted to describe the evolution of in- and out-cluster relationships over time~\cite{barat2,barat1,karsai1,Havlin,Shirazi}. These endeavors have led to uncover the heterogeneous dynamics to be a consequence of memory, e.g. in cellphone users ~\cite{2012}, face to face contacts ~\cite{barat3}, pattern of rumor spreading process ~\cite{strong,Meghdad}, individuals' web browsing pattern~\cite{barabas}, Boolean networks~\cite{Ebadi} or optimized strategy in Prisoner's Dilemma games  \cite{Lipowski,szo}.

In order to introduce the influence of memory on social networks, we have followed the Heider balance theory ~\cite{heider} that has provided a fundamental platform for tackling many sociological problems ranging from social systems ~\cite{agui,van,tylor,Leila2} to political ones ~\cite{moor}, and e.g. online networks ~\cite{esm,szell} (see ~\cite{xiao} and references therein). In this theory, relations between agents are considered as positive or negative links:  the positive sign for a link indicate for instance; friendship \cite{kun,esm}, attractiveness \cite{pat}, profit \cite{guha}, tolerance \cite{agui}, whereas a negative sign would indicate the  opposite. This theory proposes a model based on triadic configurations in which relations evolve  {in order to reduce the number of unbalanced triads and  in view of attaining "minimum tension states". Assigning a potential energy to these systems allows a more quantitative view of the landscape of the network's dynamic over time ~\cite{marvel}. 

  Let us denote  by $s_{ij}$, the link sign, corresponding to a feeling  of friendship or of animosity between the nodes $i$ and $j$.  Let the social network potential energy be calculated by summing all products of links in all triads, normalized by  the total number of triangles. From this energy point of view, minimum tension states can be distributed between global minima, namely "balanced states", and local minima, named "jammed states" ~\cite{marvel,antal,Leila1}. In presence of memory, agent relations (friendships and animosities) build up and can become vigorous over some long time. This age gain for links decreases the probability of relationship changes ~\cite{Hadis}. Gallos et al. \cite{Gall} explored the effect of nodes' attributes (e.g. age, genders) on the formation of specific configurations in real social networks. They observed that younger individuals have less triadic closure in their relations (friends of friends tie), while raising age increases this propensity.  In the spirit of scaling ideas, based on the nature of  memory features in real systems ~\cite{erthquake,neuron,Siwy}, we mimick this aging process through a kernel function which emphasizes that the strength of the relationships increases like a power law, from past to present. In so doing, we generalize the order of the derivative in the conventional continuous equation of balance theory ~\cite{equ1,equ2,equ4,Kulakoeski} in physics to a non-integer order. The results presented here below reveal the emergence of some novel states, called \emph{"glassy states"} that refer to conditions in which, for a reasonably long time interval, there is no propensity to reduce the amount of stress within the system in favor of equilibrium.

 It can be informative to compare the notions of  jammed and glassy states. Although both of these states are not   global minimum states, they have distinct characteristic that make them distinguishable. According to the tension reduction condition, complex networks tend to move toward lower energy levels and reach the global minimum states (paradise of bipolar). During this evolution a system may be trapped in some local minimum states, namely jammed states, for a while. In such a situation there is no possibility to reduce tension (energy) within a system by altering links. It has been shown that jammed states are rare events in networks dynamics and occur with negative energies. In contrast, in glassy states agents resist to modify the quality of their relationships in  order to reduce the stress which preserves the system in unstable states. Glassy states are established by a combination of aged and young links. Here, aged links refer to those that have not changed for a long time, while young links change quickly without resistance. Although there exist some links in glassy states, if changed, can guide  the system toward lower energy states, - thanks to their age the probability of change is low. It will be shown that the energy of glassy states can be negative or even positive;    due to the memory intensity, they can be the only final states of  the networks dynamical evolution.

It will be further explained that the present study considers an endogenous condition for a link switch. Exogenous shocks, or threshold of awareness ~\cite{Petroniavalanches}, are outside the present study. So we imply  a difference between simultaneous or sequential updating ~\cite{SousaEPJB66.08}. Glassy states, in condensed matter, are indeed subjects to frustration considerations and specific phase transition patterns.

\section*{The evolving network}

The dynamics of social networks is a sort of self-organized process in which relationships are modified based upon agents' benefits. These modifications occur at a microscopic level; they impose a collective behavior that guides the network  along particular structural evolution paths. In this context, Heider ~\cite{heider} proposed a "balance theory" based on triadic relations among two persons and their "attitude" towards an issue (as the third node) where connections are rearranged to become stable ~\cite{book}. According to this description, relations among individuals can be categorized into two classes based on balanced and unbalanced triadic relations. Tension, in a unbalanced triad, stimulates agents to modify their current relations to reach some balance. Later, by putting the concept of "attitude" in the background and substituting an individual as the third node, Cartwright et al.~\cite{car} generalized Heider's approach in terms of "sign graphs", in which the links among members can be $\pm 1$ (without any weight). Intuitively, a positive value denotes friendship, corporation, or tolerance, to name but a few, while a negative value depicts animosity, rivalry or intolerance in social groups~\cite{esm,kun,pat,guha}. According to the balance theory, a triadic relation among any three agents is balanced (unbalanced) only if the product of signs assigned to the links is positive (negative). Obviously, a network is "balanced" if all triangles are balanced.

Study on the dynamics of such sign networks, from the view point of balance theory, can be traced back to Antal et al. ~\cite{antal}. They proposed a model based on the evolution of links for fully connected networks, namely Constrained Triadic Dynamics $(CTD)$, which could describe how an initially imbalanced society attains a balanced state \cite{antal,Azimi}.
\begin{equation}
\frac{d}{dt} X_{ki}(t)\equiv \sum_{j=1}X_{kj}(t)X_{ji}(t).
\end{equation}
Having this $CTD$ model in hands, to investigate the memory effects on the evolution of the system, we have considered the following assumption: although changes in a relationship (link) would guide the system toward a minimum tension state, agents have less tendency to alter the nature of their connections, due to some memory of pertinent relations. This approach results in the concept of aged networks, where the history of links is associated with their ages. In previous studies, there was no discrepancy between age (duration) and strength of links in terms of weighted networks ~\cite{gra,bar,hor,gligorausloos}.

Many studies, e.g. ~\cite{Goychuk,Jeon1,west,Jeon}, have considered fractional integrals or derivatives as a generalization of ordinary differential-integral operators to non-integer ones, in view of describing the effects of past events on the present one. Such a fractional calculus approach can also be taken for the conventional balance differential equation in order to explore these effects on the social interactions.
\begin{equation}
\label{1}
^{c}_{t_0}D_{t}^{\alpha} X_{ki}(t)\equiv \sum_{j=1}X_{kj}(t)X_{ji}(t).
\end{equation}
The left hand side of Eq. ~(\ref{1}) is the Caputo ~\cite{Caputo} fractional differential operator of order $\alpha$, where $ 0 < \alpha < 1$.
This $\alpha$ "the fractional order of derivation" denotes the  significance of the memory in the interaction mechanisms: i.e., $\alpha=1$
refers to the balance theory master equation with no memory ~\cite{equ1,equ2,equ4}.

Eq. ~(\ref{1}) can be rewritten in the form of its equivalent Volterra integral ~\cite{Kilbas,Garrappa},
\begin{equation}
\label{2}
X_{ki} =X_{ki0}+\frac{1}{\Gamma{(\alpha})}\int_{t_0}^{t}dt' (t-t')^{(\alpha-1)}\left[\sum_{l=1}X_{kl}(t')X_{li}(t')\right].
\end{equation}
This functional form with a scale invariance time dependent kernel makes it possible to consider historical effects. Indeed, due to the "non-local" (more exactly, "time lag dependent") nature of these operators, past events play a "non-Markovian" role on the system dynamics. This mathematical technique practically means that those past events which  are  "further away" from the present time,
are those which less likely  contribute to the current state, - unless those incidents hold high levels of importance.

To solve the integral of Eq. ~(\ref{2}) numerically, the predictor-corrector algorithm is employed ~\cite{Garrappa,Diethelm,Lubich}. In this way, the product rectangle rule ~\cite{Garrappa} is used, in which the time domain is divided in an equispaced grid $t_j=t_0+hj$ with equal space $h$. The right hand side of Eq. ~(\ref{2}) is approximated on this grid, such that Eq. ~(\ref{2}) becomes
\begin{equation}
\label{3}
X_{ki} =X_{ki0}+ h^{\alpha}\sum_{j=0}^{n-1} b_{n-j-1} \left[\sum_{l=1}\left(X_{kl}X_{li}\right)_{j}\right].
\end{equation}
In  so doing, the $b_n=\frac{(n+1)^{\alpha}-(n)^{\alpha}}{\Gamma(\alpha+1)}$ coefficients  are the essential terms which indicate and control the role of the past  events in the model.

For the simulation part of this study, a fully connected network including $N$ nodes where all agents are acquainted with each other is first considered. Here a link between two agents $i$ and $j$ is represented by $s_{ij}$ with initial value $+1$ $(-1)$ which denotes friendship (animosity). We may start from a fully antagonistic network, where a link between any two agents has $-1$ for initial value. In order to check the evolution of the network at each time step, any link must fulfill $two$ conditions. The $first$ condition: a link is selected randomly; switching the sign of the link is made permissible, if only the total number of unbalance triangles is reduced. This description is accordance with the reduction of the total energy of the system ~\cite{marvel}, %as calculated by
\begin{eqnarray}
U=\frac{-1}{\binom{N}{3}}\sum_{i,j,k} s_{ij}s_{jk}s_{ki},
\label{e1}
\end{eqnarray}
where the sum is over all triadic relations. $U$ can accept values from $1$ (antagonistic configuration) to $-1$ (balance configuration).

The $second$ condition: a competition occurs between the tendency of a link to reduce energy and the insistence on maintaining the past relation due to the memory effect. This competitive process identifies whether the selected link (which should fulfill the condition of the $first$ step) tends to switch into the opposite sign. In this respect, a random number is generated from a normal distribution over $(0, 1)$: when this number is less than $A_{ij}^{(\alpha-1)}$, a switch into the opposite sign ($\pm 1$) is made for the selected link ($s_{ij}$). The magnitude of $A_{ij}$ is equivalent to somewhat measured by the age of the link and controls the lapse of time ("resting time") during which agents do not alter their relations. Here, $\alpha$ exhibits the rate of memory effects on relations and its value changes over $(0,1)$. Obviously, the lower the magnitude of $A_{ij}$ or the higher the value of $\alpha$ is, the more probable a link will change sign.

In the aging process, for any $N$ (the number of nodes in the system) steps, the links which do not change sign, their age ($A_{ij}$) increases by one unit. According to this dynamical evolution, a system in various paths towards minimum tension states remains unchanged  within some time intervals. When the system remains so static for  time intervals of the order of  the number of links (i.e., in order to give the same probability for each link to be chosen), we call those states \emph{glassy states}.

\section*{Results and discussions}
The dynamics of a network, in Heider balance theory, is a sort of Markovian process in which there is no evidence of the past incidents (memory) in link rearrangements, whence the evolution of relationships is entirely stochastic. Thus, in such a basic theory, when a person decides to modify a connection, apart from what happened in the past, she or he only checks the quality of a specific relation at the present moment. In terms of social networks, links carry various information such as type, age and strength of links. Age denotes the duration of a relations and strength of a link is described by its weight. Fig.  1  displays type, age and weight of links through solid (dashed) lines, colored lines and thick (thin) lines respectively for $9$ nodes which are part of a larger network. Since age of links indicates the effect of past relationships on current decisions, it can be interpreted that this parameter (age) is responsible for the memory of relationships. According to recent studies ~\cite{barat2,barat1,karsai1,Havlin}, memory imposes a scaling (increasing power law) behavior onto a system in such a way that the older the relations get, the more significant role they perform  for the network destiny. Converting the conventional continuous time equation of balance theory ~\cite{equ3,equ5} to the fractional form, Eq. (1), appropriately allows us to include and describe a memory effect over time. Thus, the interactions among individuals become time dependent and the system experiences a non-Markovian process in an endogenous way. Notice that considering the fractional space only rescales the evolution time without any impact on the phase space and the dynamics of the system. However, changes are slowing down; in other words, time intervals between changes appear to become longer; the system finally attains a balanced state later than if it is a memory-less system. %Fig. (2) illustrates such an evolution of links for a system including $16$ nodes, calculated through a numerical solution of Eq. (4) in the presence of memory ($\alpha=0.3$) and without memory ($\alpha=1$). It can be observed that the evolution of a system towards the balanced state (bipolar) is quite prolonged indeed.

%\begin{figure}
%\centerline{\includegraphics[trim = 35mm 0mm 193mm 3mm, %clip,width=8cm]{Figure-2.eps}}
%\caption{{\bf The evolution of links for a fully connected network based on Eq. (4).} These numerical solutions is considered for a network of size $16$, with ($\alpha=0.3$) and without($\alpha=1$) memory. The links initial values are taken from a uniform random distribution of numbers in the range of [-1,1]. The positive initial links are shown as red lines while negative initial links are blue. As it can be seen, a memoryless network needs less time to reach a balanced state (bipolar situation) than a network with memory.}
%\label{fig2}
%\end{figure}

In the investigation of networks dynamics based on the strict $CTD$ model, Antal et al.~\cite{antal} demonstrated that systems  may follow various paths  before obtaining the network (global) minimum tension, in a finite time interval. These systems mostly move towards final balanced states (either paradise or bipolar), where all triangles are balanced. A "paradise state" refers to a system in which all members are friendly with each other, while a "bipolar state" refers to a system which consists of $two$ main clusters, in which the members of each group are friendly with each other, but not friendly with any member of the other group. Antal et al. ~\cite{antal} also showed that there are a few "rare" states named as "jammed states" (local minima) in which a system is divided into $several$ ($\neq 2$) communities.

According to the present model, i.e., when introducing memory effects into the $CTD$ model, one even describes the formation of aged (and aging) networks. Hence, over the course of time, depending on the value of $\alpha$ which indicates the significance of relations, links can gain age with probability $A_{ij}^{(\alpha-1)}$. It can be concluded that the larger $\alpha$ is, the less relevant are effects of past events on the individual's decision. For $\alpha=1$, the memory effect completely disappears and the current model is reduced to the $CTD$ model. In the presence of memory, even though the least tension principle is enforced, the network, before obtaining either a global or a local equilibrium, can be trapped into intermediate states during several time intervals. This practically means that people do not forget their long lasting friendships or animosities readily; in other words, the system can resist toward changes over time. Fig.  2  illustrates the most prolonged periods in various paths that the system remains unchanged before reaching its balanced state, in the case of a network with $21$ nodes for $\alpha=0.7$, after $10000$ realizations. These time intervals are found to follow a Poisson distribution function $p(x)= ( 1/k!) \; e^{-\lambda}\lambda^{k} $, where $\lambda$ controls the expected frequency with which an event can occur. Such a Poisson distribution demonstrates that long time intervals become  probable; this deviation from the normal distribution leads to the emergence of inhomogeneity in the time intervals during which the system remains unchanged. Here, it is found that $\lambda$ has values $83.640$, $13.157$ for $\alpha=0.5$ and $0.7$ respectively. For larger values of $\alpha$, $\lambda$ tends to zero. In such a situation, as $\alpha$ decreases, the time intervals in which the system shows no inclination towards changes increases.

\begin{figure}
\centerline{\includegraphics[trim = 0mm 40mm 0mm 30mm, clip,width=15cm]{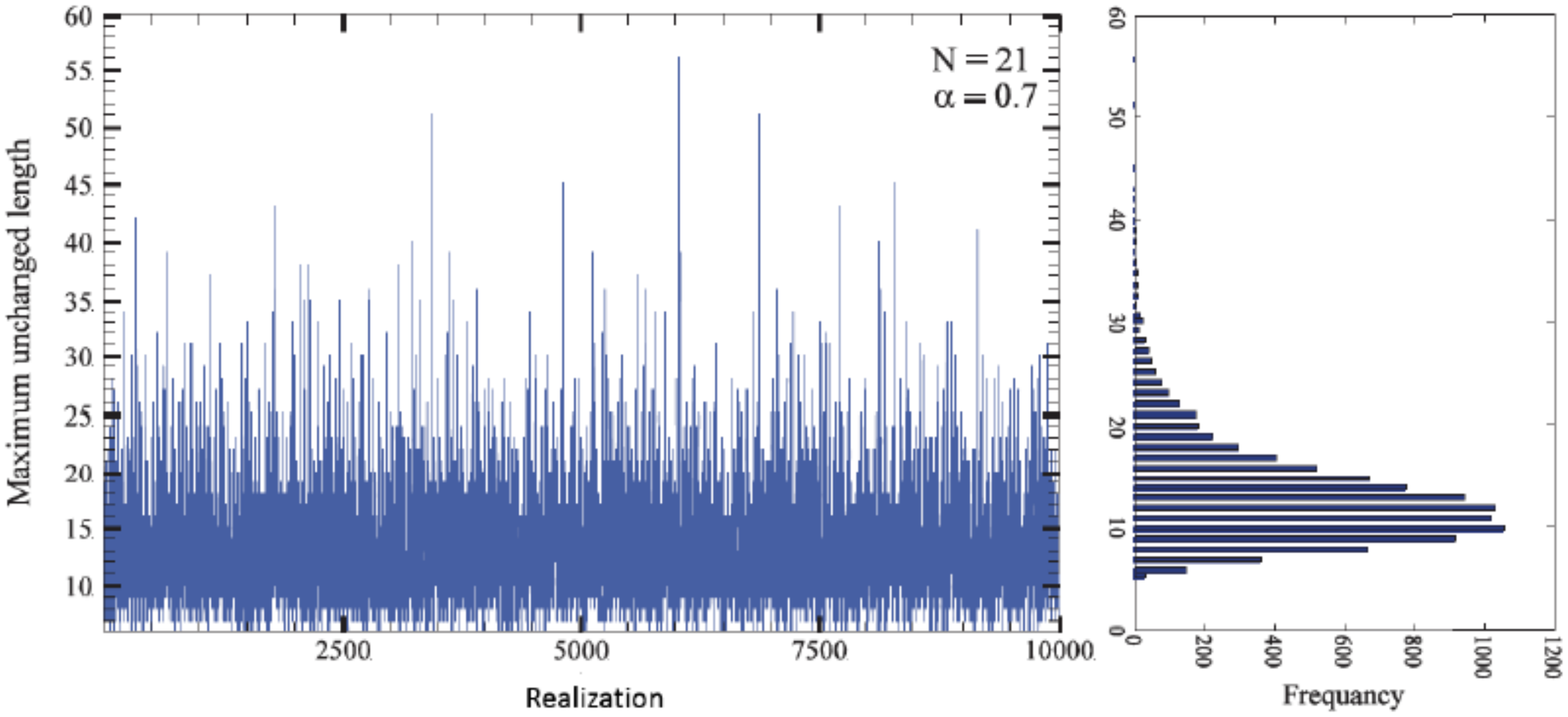}}
\caption{{\bf The maximum time interval during which a network resists against any change (long lived states).} The left panel depicts this time interval for networks with $21$ nodes in various paths (10000 realizations). The right panel illustrates that the maximum time intervals follow a Poisson distribution.}
\label{fig2}
\end{figure}

Fig.  3  illustrates such a situation: the energy of final states, Eq.~(\ref{e1}), versus the mean value of positive and negative links for a $45$ node network at different $\alpha$ values, with $10000$ realizations. It can be seen that  for $\alpha =0.3$ (left top panel) for all realizations, the system is trapped in glassy states, before it reaches its balanced states. Interestingly, however balanced and jammed states always happen in negative energies; however,  here some of glassy states occur at positive energies which imply an intrinsic instability within the system. In such cases, the system "tolerates" some high tension condition. However, no agent tends to alter the quality of her or his relationships. For larger $\alpha = 0.5$ (right top panel) or $= 0.65$ (left bottom panel in Fig. 3), the system exhibits more "flexibility", moving through the phase space; consequently, biased relationships are  later forming, implying that the system has much opportunity to reach several lower energy states with lesser tensions. In fact $\alpha$ is an indication of a time correlation within the system; the lesser the value of $\alpha$, the more the system becomes correlated to past events.
The right bottom panel in Fig.  3 displays the total number of glassy states versus the total number of
 realizations for different $\alpha$ values, 
for networks with $21$ and $45$ nodes. Accordingly, it can be seen that, when the network's members flexibility is so promoted, the chance of occupying lower energy states increases; so does the probability of obtaining balanced states. It can be practically concluded that a society containing very biased (or stubborn) people has no way to reach balanced states, but become stuck in intermediate states, namely the  long lived ($glassy$) states. On the other hand, when the intensity of commitment and/or bias in relationships decreases, a society is  enabled   to move towards balanced states in a "short" time. The right bottom panel in Fig.  3  shows a sudden change in the percentage (density) of glassy states to balanced states about $\alpha \sim 0.6$.

\begin{figure}
\centerline{\includegraphics[trim = 10mm 12mm 10mm 2mm, clip,width=.72\textwidth]{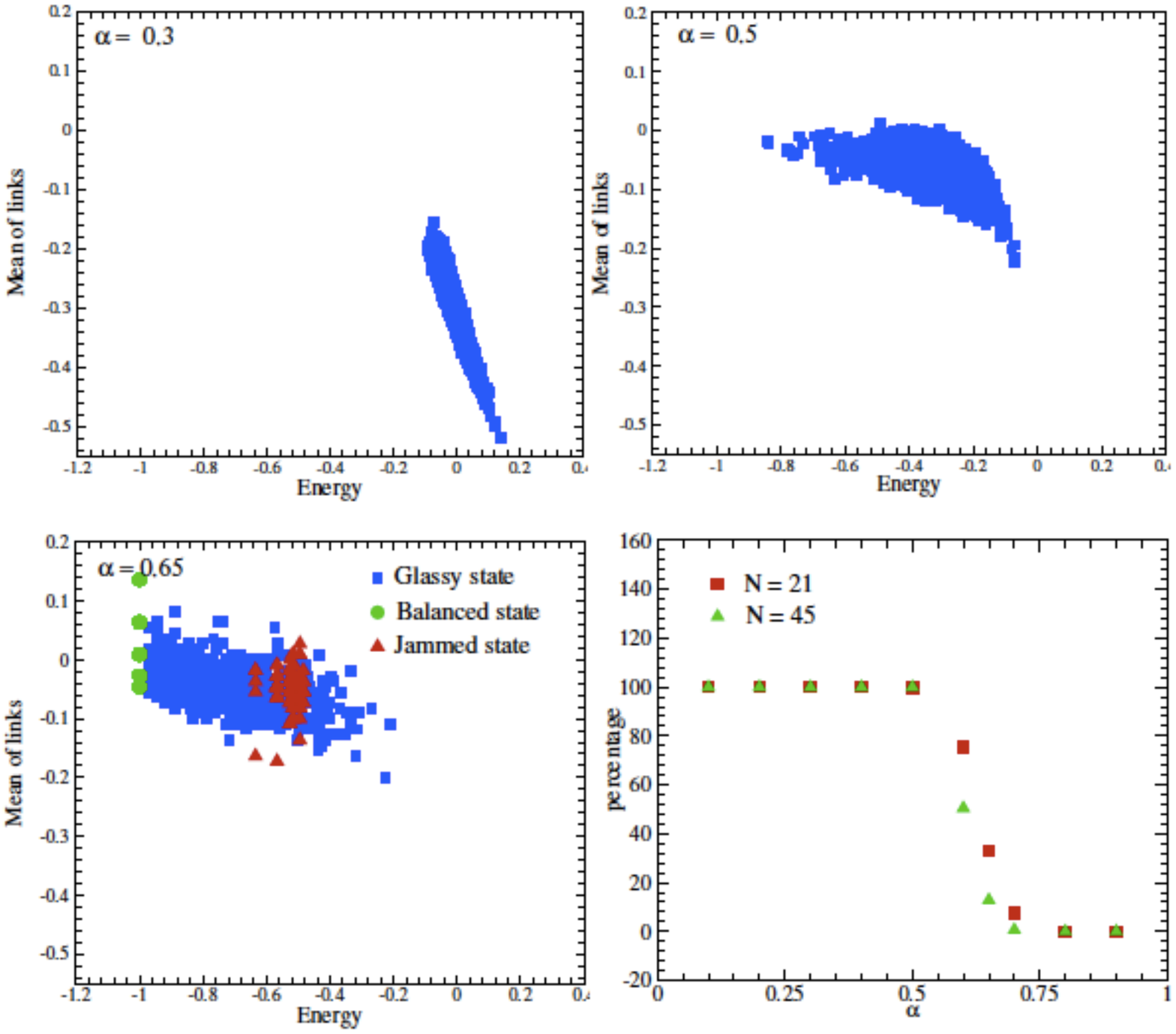}}
\caption{{\bf Energy of final states (Glassy, Balance and Jammed states) versus mean of positive and negative links.} We show these final states for networks with $45$ nodes at different $\alpha$ (strength of memory) values. With increasing $\alpha$ values, the system becomes more flexible against changes and the probability of reaching balance and jammed states increases. The right bottom panel shows a sudden "phase transition" in the percentage (density) of glassy states to balanced states about $\alpha \sim 0.6$.}
\label{fig3}
\end{figure}

\section*{Conclusions}
The memory footprint is evident in most complex systems, where life exits, from biological to economy and social systems. Thanks to the presence of memory, when we have a friendship (animosity), this relationship can get vigorous over time and the propensity of change may diminish; in other words, breaking up of aged friendship (animosity) gets difficult. Here, introducing a kernel function into the differential equation of balance theory, we provide a better insight toward considering the contribution of past events on the dynamics of a social system. We have mimicked the aging process through a kernel function which emphasizes that the disinclination of relationships to change, increases as a power law from past to present. This function leads to a fractional differential equation of order $\alpha$ where $\alpha-1$ indicates the strength of the memory effect. In other words, at high $\alpha$, links have a high probability to change and the system is more flexible toward social tension reduction.

We have investigated the feasible paths and final states that a (fully connected) network can experience, using the Constrained Triadic Dynamics(CTD) model with memory. We have employed a probability of links' disinclination to change as a $A_{ij}^{(\alpha-1)}$ function. We have found out that memory is a key factor indeed, which sometimes withstands against the quick evolution of the network and eventually preserves the system in unstable, but long lived states, namely glassy states. Under such circumstances, for various time intervals, the system has no tendency to evolve towards global or local minima. Evidences, depending on the value of the strength parameter $\alpha$, indicate that a system is enabled to reach a lower tension and energy, in non trivial path ways, visiting glassy states. In such situations, individuals become dormant or accustomed to the current state of the network. In contrast to jammed states, i.e. local minimum states, in which systems only experience negative energies, glassy states can occur in positive energy states, thereby imposing instability to and keeping stress in the system.

In conclusion, the natural concept of aging of agent based networks provides a powerful tool to investigate the dynamical
behavior of memory prone social agents, but  is also surely valid for other realistic complex systems. In real technological networks,  several links have necessarily a temporary lifetime and  fleeting; they can develop into new relations (or functions) due to  some feedback, automatic or external control, or more often reinforcement learning by memory effect over time. Therefore the concept of aged links might be a suitable language to study a wide variety of temporal networks, e.g. ranging from trading between companies, friendly and hostile relationships, scientific collaborations, political influences  through cooperation or competition,
to tweeting news in social media, beside strictly psychological aspects.

\section*{Acknowledgment}
We would like to express our gratitude to Prof. Krzysztof Kulakowski for reading the paper and for his constructive comments.
The research of GRJ was supported by the Higher Education Support Program of OSF and the Central European University.

%
%
%\nolinenumbers
%
%


\begin{thebibliography}{10}

\bibitem{heider}
Heider F.
\newblock {{A}ttitudes and cognitive organization}.
\newblock J Psychol. 1946 Dec;21:107--112.

 \bibitem{wang}
Wang Z, Szolnoki A,   Perc M.
\newblock {{S}elf-organization towards optimally interdependent networks by means of coevolution}.
\newblock New J Phys. 2014 Mar;16:033041. 

 \bibitem{perca}
Perca M, Szolnokib A.
\newblock {{C}oevolutionary gamesÑA mini review}.
\newblock BioSystems. 2010 Oct;99:109-Ð125.

\bibitem{kirman}
Kirman A, Markose S, Giansante S,  Pin P.
\newblock {{M}arginal contribution, reciprocity and equity in segregated groups: Bounded rationality and self-organization in social networks}.
\newblock  Journal of Economic Dynamics and Control. 2007 June;
31(6):2085-Ð2107.

\bibitem{Ram}
  Ramasco JJ,  Dorogovtsev SN,    Pastor-Satorras R
\newblock {{S}elf-organization of collaboration networks}.
\newblock Phys. Rev. E. 2004 Sep;70:036106. 

\bibitem{psy1}
 Becker HS.
\newblock {{N}otes on the concept of commitment}.
\newblock Am J Sociol. 1960 Jul;66(1):32--40.

\bibitem{psy2}
Stanley SM, Markman HJ.
\newblock {{A}ssessing commitment in personal relationships}.
\newblock J Marriage Fam. 1992 Aug;54(3):595--608.

\bibitem{psy3}
Clements R, Swensen CH.
\newblock {{C}ommitment to one's spouse as a predictor of marital quality among older couples}.
\newblock Curr Psychol. 2000 Jun;19(2):110--119.

\bibitem{gra}
Granovetter M.
\newblock {{T}he strength of weak ties}.
\newblock Am J Sociol. 1973 May;78(6):1360--1380.

\bibitem{barat2}
Stehl\'{e} J, Barrat A, Bianconi G.
\newblock {{D}ynamical and bursty interactions in social networks}.
\newblock Phys Rev E. 2010 March;81:035101--035104.

\bibitem{barat1}
Zhao K, Stehl\'{e} J, Bianconi G, Barrat A.
\newblock {{S}ocial network dynamics of face-to-face interactions}.
\newblock Phys Rev E. 2011 May;83:056109--056127.

\bibitem{karsai1}
Karsai M, Kaski K, Kert\'{e}sz  J.
\newblock {{C}orrelated Dynamics in Egocentric Communication Networks}.
\newblock PLoS ONE. 2012 July;7:e40612.

\bibitem{Havlin}
Rybski D, Buldyrev S, Havlin S, Liljeros F, Makse H.
\newblock {{C}ommunication activity in a social network: relation between long-term correlations and inter-event clustering}.
\newblock Scientific reports. 2012 Aug;2:560.

\bibitem{Shirazi}
Shirazi AH, Namaki A, Roohi AA, Jafari GR.
\newblock {{T}ransparency effect in emergence of monopolies in social networks}.
\newblock JASSS. 2013 Jan;6:1--10.

\bibitem{2012}
Karsai M, Kaski K, Barab\'{a}si A, Kert\'{e}sz J.
\newblock {{U}niversal features of correlated bursty behavior}.
\newblock Scientific reports. 2012 May;2:397.

\bibitem{barat3}
Vestergaard C, G\'{e}nois M, Barrat A.
\newblock {{H}ow memory generates heterogeneous dynamics in temporal networks}.
\newblock Phys Rev Lett. 2014 Oct;90:042805.

\bibitem{strong}
Karsai M, Perra N, Vespignani A.
\newblock {{T}ime varying networks and the weakness of strong ties}.
\newblock Scientific reports. 2014 Feb;4:4001--4007.

\bibitem{Meghdad} Saeedian M, Khaliqi M, Azimi-Tafreshi N, Jafari GR, Ausloos M.
\newblock {{M}emory effects on epidemic evolution: The susceptible-infected-recovered epidemic model}.
\newblock  Phys. Rev. E 2017 Feb;95(2):022409

\bibitem{barabas}
Dezso Z, Almaas E, Lukacs A, Racz B, Szakadat I, Barab\'{a}si A.
\newblock {{D}ynamics of information access on the web}.
\newblock Phys Rev E. 2008 Jun;73: 066132--066137.

\bibitem{Ebadi}
Ebadi E, Saeedian M, Ausloos M, Jafari GR.
\newblock {{E}ffect of memory in non-Markovian Boolean networks illustrated with a case study: A cell cycling process}.
\newblock EPL (Europhysics Letters) 2016 Des;116(3): 30004.

\bibitem{Lipowski}
Lipowski A, Gontarek K, Ausloos M.
\newblock {{S}tatistical mechanics approach to a reinforcement learning model with memory}.
\newblock Physica A. 2009 May;388:1849--1856.

 \bibitem{szo}
Szolnoki A, Perc M, Szab$\acute{o}$ G,   Stark H-U.
\newblock {{I}mpact of aging on the evolution of cooperation in the spatial prisonerÕs dilemma game}.
\newblock Phys Rev E. 2009 Aug;80:021901.

\bibitem{agui}
Aguiar F, Parravano A.
\newblock {{T}olerating the Intolerant: Homophily, Intolerance, and Segregation in Social Balanced Networks}.
\newblock J Conflict Resolut. 2013 Aug;59(1):29--50.

\bibitem{van}
van de Rijt A.
\newblock {{T}he Micro-Macro Link for the Theory of Structural Balance}.
\newblock J Math Sociol. 2011 Feb;35:94--113.

\bibitem{tylor}
Summers TH, Shames I.
\newblock {{A}ctive influence in dynamical models of structural balance in social networks}.
\newblock Europhys Lett. 2013 Jul;103:18001.

\bibitem{Leila2}
Hassanibesheli F, Hedayatifar L, Gawro{\'n}ski P, Stojkow M, {\.Z}uchowska-Skiba D, Ku?akowski K.
\newblock {{G}ain and loss of esteem, direct reciprocity and Heider balance}.
\newblock Physica A. 2017 Feb;468:334-339.

\bibitem{moor}
Moore M.
\newblock {{A}n international application of Heider's balance theory}.
\newblock Eur J Soc Psychol. 1978 Jul;8:401--405;
Moore M.
\newblock {{S}tructural balance and international relations}.
\newblock Eur J Soc Psychol. 1979 Sep;9:323--326.

 \bibitem{esm}
Esmailian P, Abtahi SE, Jalili M.
\newblock {{M}esoscopic analysis of online social networks: The role of negative ties}.
\newblock Phys Rev E. 2014 Oct;90:042817.

\bibitem{kun}
Kunegis J, Lommatzsch A and Bauckhage C.
\newblock {{T}he slashdot zoo: Mining a social network with negative edges}. in
\newblock Proceedings of the 18th international conference on World wide web - WWW 2009 (Assoc Comput Machinery, New York), 2009 ;741--750.

\bibitem{pat}
Doreian P.
\newblock {{E}volution of Human Signed Networks}.
\newblock Metodoloski zvezki. 2004; 1(2):277--293.

\bibitem{guha}
Guha RV, Kumar R, Raghavan P, Tomkins A.
\newblock {{P}ropagation of trust and distrust}. in
\newblock Proceedings of the 13th international conference on World Wide Web, 2004 May 17-20, New York, NY, USA.

\bibitem{szell}
Szell M, Lambiotte R, Thurner S.
\newblock {{M}ultirelational organization of large-scale social networks in an online world}.
\newblock Proc Natl Acad Sci USA. 2010 Jun;107:13636.

\bibitem{xiao}
Zheng X, Zeng D, Wang FY.
\newblock {{S}ocial balance in signed networks}.
\newblock Inf Syst Front. 2014 Jan;17(5):1--19.

\bibitem{marvel}
Marvel SA, Kleinberg J, Strogatz SH.
\newblock {{T}he energy landscape of social balance}.
\newblock Phys Rev Lett. 2009 Nov;103:198701.

\bibitem{antal}
Antal T, Krapivsky P, Redner S.
\newblock {{S}ocial balance on networks: The dynamics of friendship and enmity}.
\newblock Physica D. 2006 Dec;224:130--136.

\bibitem{Leila1}
Hedayatifar L, Hassanibesheli F, Shirazi AH, Vasheghani Farahani S,  Jafari GR
\newblock{{P}seudo paths toward minimum energy states in network dynamics.}
\newblock Physica A. 2017 Apr; doi.org/10.1016/j.physa.2017.04.132.

\bibitem{Hadis}
Safdari H, Chechkin AV, Jafari GR, Metzler R.
\newblock {{A}ging Scaled Brownian Motion}.
\newblock Phys Rev E. 2015 Apr;91:042107.

 \bibitem{Gall}
Gallos L K, Rybski D, Liljeros F, Havlin S,   Makse H A.
\newblock {{H}ow People Interact in Evolving Online Affiliation Networks}.
Phys Rev X. 2012 Aug;2:031014.
 

\bibitem{erthquake}
Livina V, Havlin S, Bunde A.
\newblock {{M}emory in the Occurrence of Earthquakes}.
\newblock Phys Rev Lett. 2005 Nov;95:208501.

 \bibitem{neuron}
 Kemuriyama T, Ohta H, Sato Y, Maruyama S, Tandai-Hiruma M.
\newblock {{A} power-law distribution of inter-spike intervals in renal sympathetic nerve activity in salt-sensitive hypertension-induced chronic heart failure}.
\newblock BioSystems. 2010 Agu; 101:144--147.

 \bibitem{Siwy}
 Siwy Z, Ausloos M,  Ivanova K.
\newblock {{C}orrelation studies of open and closed states fluctuations in an ion channel: Analysis of ion current through a large-conductance locust potassium channel}.
\newblock Phys Rev E. 2002 Feb; 65:031907.

\bibitem{Kulakoeski}
Kulakoeski K, Gawronski P,   Gronek P.
\newblock {{E}pidemic spreading on evolving signed networks}.
\newblock Int. J. Mod. Phys. C 2005 Nov:16;707.

\bibitem{equ1}
Ku{\l}akowski K, Gawronski P, Gronek P.
\newblock {{T}he Heider balance-a continuous approach}.
\newblock Int J Mod Phys C. 2005; 16:707--716.

\bibitem{equ2}
Marvel SA, Kleinberg J, Kleinberg RD and Strogatz SH.
\newblock {{C}ontinuous-time model of structural balance}.
\newblock Proc Natl Acad Sci USA. 2011 Feb; 108:1771--1776.

\bibitem{equ4}
Altafini C.
\newblock {{D}ynamics of Opinion Forming in Structurally Balanced Social Networks}.
\newblock PLoS ONE. 2012 Jun; 7:38135.

\bibitem{Petroniavalanches}
Ausloos M, Petroni F.
\newblock {{T}hreshold Model for Triggered Avalanches on Networks}. in
\newblock Stock Markets, F. Petroni, F. Prattico and G. D'Amico. Eds. (Nova Sc., New York) 2013 ;83--101.

\bibitem{SousaEPJB66.08}
Sousa AO, Yu-Song T, Ausloos M.
\newblock {{P}ropaganda spreading or running away from frustration effects in Sznajd model}.
\newblock Eur Phys J B. 2008 Oct; 66:115--124.

\bibitem{book}
Newcomb TM, Turner RH, Converse PE.
\newblock {{S}ocial Psychology: The Study of Human Interaction}.
\newblock (New York: Holt, Rinehart and Winston). %1965

\bibitem{car}
Cartwright D,  Harary F.
\newblock {{S}tructure balance: A generalization of Heider's theory}.
\newblock Psychol Rev. 1956 Sep; 63:277--293.
% (1956)

\bibitem{Azimi}
Saeedian M, Azimi-Tafreshi N, Jafari GR, Kertesz J.
\newblock {{E}pidemic spreading on evolving signed networks}.
\newblock Phys Rev E. 2017 Apr;95:022314.

\bibitem{bar}
Barrat A, Barthelemy M, Pastor-Satorras R, Vespignani A.
\newblock {{T}he architecture of complex weighted networks}.
\newblock PNAS. 2004 Mar; 101:3747--3752.

\bibitem{hor}
Horvath S.
\newblock {{W}eighted Network Analysis. Applications in Genomics and Systems Biology}.
\newblock Springer Book. 2011 %ISBN 978-1-4419-8818-8.

\bibitem{gligorausloos}
Gligor M, Ausloos M.
\newblock {{C}lusters in weighted macroeconomic networks: the EU case. Introducing the overlapping index of  GDP/capita fluctuation correlations}.
\newblock Eur Phys J B. 2008 Jun; 63:533--539.

\bibitem{Goychuk}
Goychuk I.
\newblock {{V}iscoelastic subdiffusion: From anomalous to normal}.
\newblock Phys Rev E. 2009 Oct; 80:046125.

\bibitem{Jeon1}
Jeon JH, Metzler R.
\newblock {{F}ractional Brownian motion and motion governed by the fractional Langevin equation in confined geometries}.
\newblock Phys Rev E. 2010 Feb; 81:021103.

\bibitem{west}
West B J, Turalska M, Grigolini P.
\newblock {{F}ractional calculus ties the microscopic and macroscopic scales of complex network dynamics}.
\newblock New J Phys. 2015 Apr; 17:045009.

\bibitem{Jeon}
Safdari H, Kamali MZ, Shirazi AH, Khaliqi M, Jafari GR, Ausloos M.
\newblock {{F}ractional Dynamics of Network Growth Constrained by Aging Node Interactions}.
\newblock PloS ONE 11 (5), e0154983 2016.

\bibitem{Caputo}
Caputo M.
\newblock {{L}inear Models of Dissipation whose Q is almost Frequency Independent-II}.
\newblock Geophys. J. R. Astron. Soc. 1967 May; 13:529.

\bibitem{Kilbas}
Kilbas AA, Srivastava HM, and Trujillo JJ.
\newblock {{T}heory and Applications of Fractional Differential Equations}. (North-Holland Mathematics Studies, Elsevier Science, Amsterdam) 2006.

\bibitem{Garrappa}
Garrappa R.
\newblock {{O}n linear stability of predictor-corrector algorithms for fractional differential equations}.
\newblock Int J Comp Math. 2010 Aug; 87(10):2281--2290.

\bibitem{Diethelm}
Diethelm K, Freed AD.
\newblock {{T}he FracPECE subroutine for the numerical solution of differential equations of fractional order}.
\newblock In Forschung und Wissenschaftliches Rechnen. 1999, eds. Heinzel S, and Plesser T, Gessellschaft f\"{u}r wissenschaftliche Datenverarbeitung, G\"{o}ttingen, 1998 ;57--71.

\bibitem{Lubich}
Lubich C.
\newblock {{A} stability analysis of convolution quadratures for Abel-Volterra integral equations}.
\newblock IMA J Numer Anal. 1986; 6:87--101.

\bibitem{equ3}
Traag V, Van Dooren P and De Leenheer P.
\newblock {{D}ynamical Models Explaining Social Balance and Evolution of Cooperation}.
\newblock PLoS ONE. 2013 Apr; 8:60063.

\bibitem{equ5}
Krawczyk MJ, Castillo-Mussot M, Hern\'{a}ndez-Ram\'{\i}rez E, Naumis GG, Ku{\l}akowski K.
\newblock {{H}eider balance, asymmetric ties, and gender segregation}.
\newblock Physica A. 2015 Dec; 439:66--74.

%\bibitem{DrozdzFENS}
%Drozsz S.
%\newblock {{C}omplexity characteristics of world  econo- and sociophysics scientific collaboration network}.
%\newblock Acta Phys. Pol. A in press 2016.


\end{thebibliography}
\end{document}